\documentclass[conference]{IEEEtran}
\IEEEoverridecommandlockouts
\usepackage{cite}
\usepackage{amsmath,amssymb,amsfonts}
\usepackage{algorithmic}
\usepackage{graphicx}
\usepackage{textcomp}
\usepackage{xcolor}
\usepackage{url}
\usepackage{tabularx}
\usepackage{booktabs}
\usepackage{microtype} 
\graphicspath{{images/}{./}}

\def\BibTeX{{\rm B\kern-.05em{\sc i\kern-.025em b}\kern-.08em
    T\kern-.1667em\lower.7ex\hbox{E}\kern-.125emX}}

\usepackage{enumitem}
\setlist[itemize]{noitemsep, topsep=2pt, parsep=0pt, partopsep=0pt}
\setlist[enumerate]{noitemsep, topsep=2pt, parsep=0pt, partopsep=0pt}

\usepackage[font=small,labelfont=bf]{caption}

\begin{document}

\title{Hybrid Workflow Composition for Extreme-Scale Data Processing: A Case Study on the HL-LHC (Extended Version)}

\author{\IEEEauthorblockN{Alan Malta Rodrigues\thanks{A. Malta Rodrigues is a researcher with the Department of Physics, University of Notre Dame. He serves as the Lead Developer for the CMS Workload Management System; the findings in this study are intended to guide the architectural evolution of the next-generation CMS orchestration software.}}
\IEEEauthorblockA{\textit{Department of Computer Science and Engineering} \\
\textit{University of Notre Dame}\\
Notre Dame, IN \\
amaltar2@nd.edu}
\and
\IEEEauthorblockN{Douglas Thain}
\IEEEauthorblockA{\textit{Department of Computer Science and Engineering} \\
\textit{University of Notre Dame}\\
Notre Dame, IN \\
dthain@nd.edu}
}

\maketitle

\begin{abstract}
High-Throughput Computing (HTC) environments tailored for high-concurrency resource efficiency require sophisticated orchestration to manage petabyte-scale data across heterogeneous resources. A critical but often overlooked challenge is workflow composition: the strategic grouping of tasksets within a Directed Acyclic Graph (DAG) to mitigate execution overhead while maximizing resource utilization. This paper presents a novel simulation framework for characterizing the interplay between taskset granularity and system-level constraints (e.g., job latency, failure rate, throughput, and I/O bandwidth).

By exploring a high-dimensional parameter space, we quantify the performance sensitivity of diverse workflow topologies. Our results demonstrate that hybrid composition strategies, which dynamically balance taskset independence with execution grouping, can yield up to $3.8\times$ throughput increase and a $14.9\times$ reduction in network overhead. We further propose a multi-metric objective function that enables policy-driven optimization, allowing system architects to navigate the Pareto frontier between throughput, I/O cost, and CPU efficiency. These findings provide a rigorous foundation for automated workflow synthesis in distributed systems, offering a scalable model for next-generation scientific pipelines. All artifacts are publicly available.
\end{abstract}


\section{Introduction}
\label{sec:intro}
Workflow construction and execution strategies have a significant impact on resource utilization, event throughput, and operational efficiency in large-scale scientific computing. As data-intensive domains transition toward exascale processing, the interplay between workflow topology (the logical grouping of execution stages) and heterogeneous resource constraints becomes a primary performance bottleneck.

In High-Energy Physics (HEP), processing pipelines operate on \textbf{events}: atomic physics collision records generated by particle detectors. Workflows are modeled as Directed Acyclic Graphs (DAGs) where nodes represent \textbf{tasksets} (distinct computational steps, such as detector simulation or track reconstruction). During execution, these tasksets \textbf{expand}, meaning they are instantiated and split across event batches into hundreds of thousands of schedulable grid jobs that execute across distributed sites. While batch schedulers determine when and where jobs run across grid sites, initial workflow composition determines what is packaged inside each job. Grouping dependent tasksets into a single job enables direct data exchange via local node storage (scratch disk), whereas executing them as separate jobs forces intermediate data to be written to and read from remote grid storage.

The Compact Muon Solenoid (CMS) \cite{ref1} experiment at the Large Hadron Collider (LHC) \cite{ref2} typifies this challenge, generating hundreds of petabytes of data annually. To process this volume, HEP currently relies on the Worldwide LHC Computing Grid (WLCG) \cite{ref3}, a global federation of 1.4 million cores. However, as processing demands increase for the High-Luminosity LHC (HL-LHC) era \cite{ref4}, workflow orchestration must adapt to leverage increasingly heterogeneous opportunistic, cloud and HPC resources alongside traditional grid sites.

The CMS workflow management system has evolved over the past 15 years of LHC operations, primarily providing two static execution models: \textbf{TaskChain}, which executes each taskset as an independent, standalone grid job, and \textbf{StepChain}, which merges all sequential tasksets into a single monolithic job running on a worker node. While effective, these represent two extremes of a vast, unexplored compositional spectrum. There exists a significant research gap in exploring adaptive, group-based compositions that could better leverage the increasingly heterogeneous set of opportunistic, cloud, and High-Performance Computing (HPC) resources.

To systematically evaluate these compositional trade-offs, we investigate the following \textbf{Research Questions (RQs)}:
\begin{itemize}
    \item \textbf{RQ1}: How can dependent tasksets be grouped into execution units that respect resource constraints while optimizing event throughput?
    \item \textbf{RQ2}: What are the fundamental trade-offs between I/O activity, resource utilization, and throughput across the grouping spectrum?
    \item \textbf{RQ3}: What is the sensitivity of this composition strategy to target job length, failure rate, network bandwidth, and workflow profiles?
\end{itemize}

To address this gap, this paper introduces a framework for the design, simulation, and evaluation of group-based DAG workflow strategies. Our \textbf{primary contributions} include:
\begin{itemize}
    \item \textbf{Workflow Composition Engine}: A formalization and framework that parses generic workflows and enumerates all valid group-based compositions based on taskset dependencies and hard constraints.
    \item \textbf{DAGFlowSim Engine}: A high-fidelity simulation engine \cite{ref5} capable of executing thousands of compositions under parameterized scenarios, including variable failure rates, job length and client-side network bandwidth.
    \item \textbf{Multi-Metric Analytic Framework}: A quantitative characterization of the trade-offs between throughput and efficiency, providing a policy-driven score function for strategy selection.
    \item \textbf{Open-Source Artifacts}: We provide the simulator, analysis scripts, and a dataset covering thousands of simulated compositions to support community reproducibility.
\end{itemize}

\begin{figure}[t]
\centering
\includegraphics[width=\columnwidth]{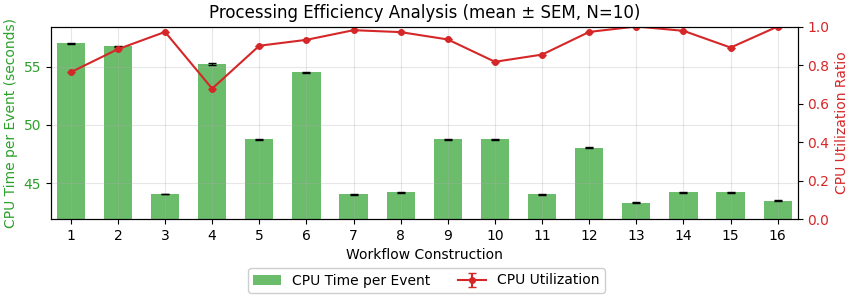}
\caption{Execution efficiency across 16 workflow constructions for a 5-taskset realistic sequential workflow (12h target job length, 5\% failure rate, 100~MB/s network bandwidth). Green bars and red points represent mean values across $N=10$ independent simulation runs with randomized failure seeds. Vertical whiskers indicate the Standard Error of the Mean ($\text{SEM} \approx 0.1\%$). Constructions 1 (StepChain) and 16 (TaskChain) map existing production paradigms, exposing critical trade-offs between resource utilization and processing overhead.}
\label{fig:introduction_processing_analysis}
\end{figure}

Fig.~\ref{fig:introduction_processing_analysis} illustrates the utility of this framework, showing processing efficiency metrics across 16 workflow constructions evaluated over $N=10$ independent simulation runs per construction. By exposing the spectrum between existing extrema (represented by constructions 1 and 16), our framework enables the identification of optimal intermediate strategies. To ensure the reliability of these simulated results, we performed a validation study using real CMS workflow schemas, demonstrating that the simulator accurately reflects the trade-offs observed in production environments.

\section{Background and Related Work}
\label{sec:background}

Efficiently executing petabyte-scale scientific workflows requires integrating Directed Acyclic Graph (DAG) orchestration, resilient High-Throughput Computing (HTC), and high-fidelity simulation. This section situates \textit{DAGFlowSim} within the broader landscape of extreme-scale computing middleware.

\subsection{The CMS Processing Model as an Extreme-Scale Use Case}

\begin{figure}[t]
\centering
\includegraphics[width=\columnwidth]{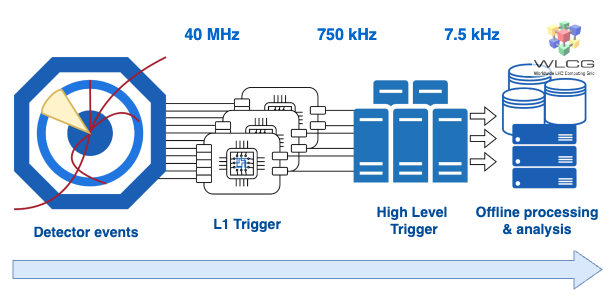}
\caption{The CMS Data Processing Pipeline, illustrating the data flow from the online detector triggers to prompt reconstruction and archival storage. Projected HL-LHC data rates (750~kHz L1 and 10~kHz HLT) represent a $10\times$ increase over current operations, necessitating the optimized offline processing and simulation orchestration evaluated in this study.}
\label{fig:cms-pipeline}
\end{figure}

The CMS experiment at the LHC generates data at an unprecedented scale, necessitating a multi-stage filtering and processing chain (Fig.~\ref{fig:cms-pipeline}). As the field transitions to the High-Luminosity LHC (HL-LHC) era, the system must evolve to handle a significant increase in data throughput. While current operations manage a Level-1 trigger rate of 100~kHz and a High-Level Trigger (HLT) output of 1~kHz, the HL-LHC is projected to increase these rates to 750~kHz and 10~kHz, respectively \cite{ref15, ref18}.

From a distributed systems perspective, this nearly ten-fold increase in event rates - coupled with rising event complexity - represents a massive orchestration challenge. The \textit{Offline Processing} stage must transform raw data into physics-ready formats through complex reconstruction and simulation DAGs. These workflows are executed across the Worldwide LHC Computing Grid (WLCG), a heterogeneous infrastructure where system performance is increasingly dominated by I/O constraints and the structural efficiency of the workload rather than peak floating-point performance.

\subsection{Workflow Orchestration and Composition}

Scientific workflows are typically modeled as DAGs, where nodes represent computational tasksets and edges denote data dependencies. In High-Energy Physics (HEP), Workflow Management Systems (WMS) such as WMCore \cite{ref10}, PanDA \cite{ref11}, and DIRAC \cite{ref12} serve as middleware layers that decouple logical workflow definitions from physical execution units. 

Recent characterizations of extreme-scale WMS \cite{ref19, ref21} identify a critical design gap: the transition from logical DAGs to schedulable execution units. While traditional tools like Pegasus \cite{ref8} and DAGMan \cite{ref7} manage job-level dependencies, the \textit{composition strategy} - the logic used to aggregate tasks into jobs - remains a static policy. Thomas and Thain \cite{ref6} established that dependency-based grouping improves data locality; however, as workloads become increasingly data-plane driven \cite{ref17} and cross-facility in nature \cite{ref20}, the need for scalable composition techniques that can navigate the trade-off between taskset granularity and I/O pressure becomes paramount \cite{ref16}. \textit{DAGFlowSim} addresses this by enabling the systematic exploration of the composition spectrum.

\subsection{HTC Resilience and Heterogeneity}
Unlike High-Performance Computing (HPC), HTC focuses on aggregate throughput over long-running campaigns. HEP workloads are predominantly HTC-based, relying on batch schedulers (e.g., HTCondor, Slurm) to match millions of jobs to distributed, often opportunistic resources. Resilience in these environments is achieved through retry policies to mitigate hardware and network faults. Since failure rates directly inflate wall-clock time and resource waste, \textit{DAGFlowSim} incorporates configurable failure models to study how different taskset composition patterns (e.g. monolithic vs granular) impact global system resilience and throughput.

\subsection{Simulation and Modeling Approaches}

Simulation is essential for evaluating composition policies without the prohibitive cost of full-scale grid experiments. While robust simulation frameworks like WRENCH \cite{ref22} (built on SimGrid) excel at high-fidelity network topology and queuing emulation, their setup complexity and execution overhead scale significantly when conducting high-dimensional combinatorial parameter sweeps.

\textit{DAGFlowSim} complements these established frameworks by providing a lightweight, specialized discrete-event engine optimized specifically for the structural analysis of HEP-style workloads. By decoupling heavy network-state simulation from taskset aggregation dynamics, DAGFlowSim makes evaluating thousands of multi-variate scenarios (e.g., 7,680 simulation runs) computationally tractable. Specifically, it differentiates itself from general-purpose simulators by emphasizing: (1) event-based taskset scaling; (2) group-based scheduling and execution; (3) systematic sensitivity analysis across job lengths, failure rates, and bandwidth constraints; and (4) a specialized multi-metric analytic framework for characterizing the Pareto frontier of workflow composition and execution.

\section{System Model and DAGFlowSim Architecture}
\label{sec:architecture}
\begin{figure}[t]
\centering
\includegraphics[width=\columnwidth]{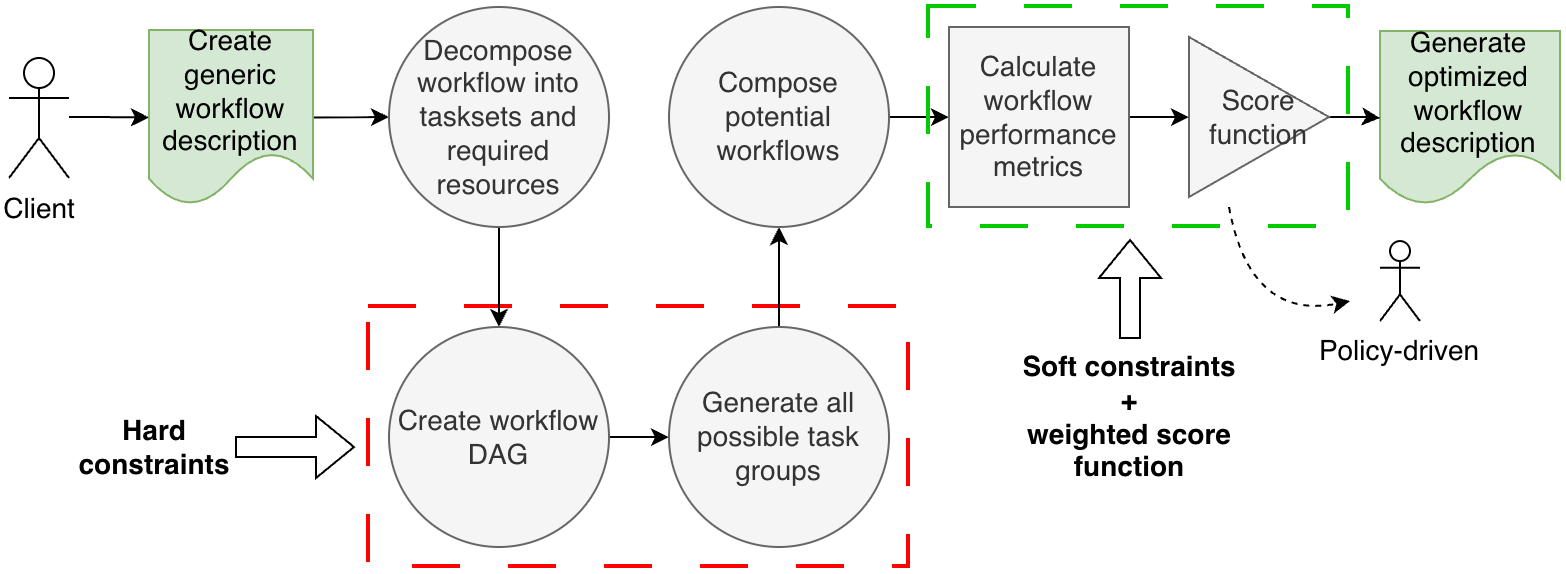}
\caption{System architecture of the proposed orchestration framework, illustrating the pipeline from generic workflow specification to optimized policy-driven compositional ranking.}
\label{fig:construction-flow}
\end{figure}

The primary objective of the proposed framework is to transform an abstract DAG workflow description into an optimized, group-based execution plan tailored to specific infrastructure policies. The system architecture, illustrated in Fig.~\ref{fig:construction-flow}, integrates taskset grouping logic with a high-fidelity simulation engine to navigate the compositional trade-off space.

\subsection{Design Goals and Policy-Driven Selection}
\label{sec:design}

The framework satisfies four primary requirements: (1) automated taskset grouping that enforces hard constraints; (2) discrete-event simulation of group-based execution; (3) systematic composition search space exploration; and (4) a flexible, multi-objective score function.

Here, \textbf{hard constraints} represent parameters strictly necessary for the taskset grouping decision, such as DAG dependency ordering, target OS version, and CPU architecture. In contrast, \textbf{soft constraints} represent parameters that do not require strict compatibility, such as memory and CPU cores required for taskset execution, input/output data rates, and other variables that influence operational optimization objectives and site policies - such as targeting ideal job durations (e.g., 8--12 hours) or capping remote network bandwidth usage.

\subsection{Assumptions and Scope}
\label{sec:assumptions}
To maintain a focused and computationally tractable simulation environment, the following assumptions define the model's analytical scope:

\begin{itemize}
    \item \textbf{Deterministic Taskset Characterization}: Tasksets are defined by CPU time and data size per event, assuming uniform CPU performance across worker nodes. \textit{Rationale}: This isolates the structural impact of workflow composition from hardware noise and node heterogeneity (analytical isolation).
    \item \textbf{Operational Overhead}: The model incorporates taskset bootstrap costs (60s) and client-side network bandwidth for remote I/O. \textit{Rationale}: Baseline overhead values are grounded in production WLCG execution logs \cite{ref10}, while high-level WMS bookkeeping and site-level storage queuing latencies are excluded to isolate job-level composition dynamics.
    \item \textbf{Structural Constraints}: The framework supports Directed Acyclic Graphs (DAGs), including sequential chains and fork structures. \textit{Rationale}: For this baseline study, tasksets are assumed to be software-compatible, allowing any grouping configuration that respects DAG dependency order; hardware-incompatible tasksets are enforced via hard constraints in the compatibility matrix $H$.
    \item \textbf{Comparative Analysis}: DAGFlowSim is designed for the relative ranking of compositions rather than the absolute prediction of production metrics. \textit{Rationale}: This enables researchers to evaluate and identify optimal compositional strategies for a given scenario without requiring an exhaustive, high-overhead model of every transient grid anomaly.
    \item \textbf{Workflow Granularity and Scale}: In HTC environments typified by CMS, extreme scale derives from event concurrency and petabyte-scale data volume rather than DAG topological depth. Standard production workflows consist of compact sequential or fork structures ($N \le 10$ processing stages, e.g., GEN-SIM-DIGI-RECO-NANOAOD). For these representative lengths, complete compositional enumeration is computationally trivial (16 valid sequential partitions for $N=5$). For hypothetical ultra-large DAGs ($N > 10$), partition scaling would require heuristic search techniques left for future work.
\end{itemize}

\subsection{Mathematical Formulation of Taskset Grouping}
\label{sec:algebraic}

We formalize the workflow composition as a constrained partitioning problem. Let $W = \{T_1, T_2, \ldots, T_n\}$ denote the set of workflow tasksets. A \textbf{compatibility matrix} $H$ is defined such that $H_{ij} = 1$ if $T_i$ and $T_j$ satisfy hardware homogeneity (CPU/OS) and share a dependency path in the workflow DAG.

\textbf{Definition 1 (Valid Group):} A subset of tasksets $G \subseteq W$ is a valid execution group if it satisfies:
\begin{enumerate}
\item \textit{Pairwise Compatibility}: $\forall T_i, T_j \in G,\; H_{ij} = 1$.
\item \textit{Path Containment}: For any $T_i, T_j \in G$, every taskset $T_k$ on the directed path between $T_i$ and $T_j$ in the DAG must also satisfy $T_k \in G$.
\end{enumerate}

The path containment requirement ensures that groups do not skip intermediate dependencies, preserving execution order within a single atomic execution unit (a grid job).

\textbf{Definition 2 (Workflow Composition):} A composition $\mathcal{C}$ is a partition of $W$ into disjoint groups $\{G_1, G_2, \ldots, G_n\}$ such that $G_k \in \mathcal{G}$ for all $k$, and $\bigcup_{k=1}^m G_k = W$.

The workflow builder enumerates the space of all valid compositions $\mathcal{C}$, ranging from the fully-ungrouped \textit{TaskChain ($m = n$)} to the fully-grouped \textit{StepChain ($m = 1$)}.

\subsection{Workflow Model and Taskset Properties}
\label{sec:workflow}

We modeled the workflow as an annotated DAG where each node $T_i$ represents a taskset of idempotent jobs and edges represent data dependencies. Unlike job-level simulators, our model enriches the taskset definition with metadata required for structural orchestration and data-locality optimization.

\begin{itemize}
    \item \textbf{Taskset Characteristics}: Each taskset is defined by its computational intensity (CPU time per event) and I/O footprint (data size per event). Resource requirements - including memory, core count, OS version and CPU architecture constraints - are treated as invariants that dictate group compatibility. A \textit{KeepOutput} flag designates whether a taskset's data is transient (local to the group) or permanent (requiring remote storage write-back), unless data dependency is across group boundaries.
    \item \textbf{Group-Based Execution}: A group $G$ represents the unit of grid scheduling. Tasksets within a group execute sequentially on the same worker node, enabling intra-job data locality where intermediate data is passed via local scratch space rather than shared storage. This localized I/O mitigates network pressure and drives the performance gains evaluated in Section~IV.
\end{itemize}

\subsection{Simulation Engine}
\label{sec:simulation}
DAGFlowSim employs a discrete-event, batch-synchronous simulation model to characterize workflow performance under varying system pressures. The execution logic is governed by three primary mechanisms.

\textbf{Job Sizing and Events}: The total workload is defined by a request for $N$ events. The simulator calculates the batch size (events per job) by dividing the target job length by the aggregate $TimePerEvent$ of all tasksets within a group. This ensures that every grid job instance within a group processes a uniform number of events, optimizing for typical site-specific time limits (e.g., 8-hour or 24-hour slots).

\textbf{Event-Driven Flow}: A group enters the \textit{Ready} state only when its parent dependencies in the DAG are satisfied. We model a steady-state throughput environment where jobs are dispatched in batches. Upon completion, processed events propagate to child groups. This flow continues until all $N$ events reach the DAG leaf nodes.

\textbf{Overhead and Resilience}: Each job's wall-clock time includes a fixed bootstrap overhead (60s per taskset) and a variable I/O latency determined by $SizePerEvent$ and a configurable network bandwidth (default 100 MB/s). A probabilistic failure model forces failed jobs to consume partial resources before returning their events to the group buffer for retry, enabling robust characterization under grid instability.

\subsection{Performance Metrics}
\label{sec:metrics}
To enable a multi-dimensional comparison of compositions, the framework calculates job-level performance metrics and aggregate them into workflow-level metrics. Table~\ref{tab:metrics} categorizes these metrics.

\begin{table}[ht]
\centering
\caption{Taxonomy of Workflow and Job-Level Metrics.}
\label{tab:metrics}
\small 
\begin{tabularx}{\columnwidth}{@{}l X@{}}
\toprule
\textbf{Category} & \textbf{Metric (Description)} \\
\midrule
\textbf{Throughput} & \textbf{Event Throughput}: Normalized events per CPU-second. \\
\midrule
\textbf{Latency} & \textbf{Total Turnaround}: End-to-end workflow makespan. \\
                    & \textbf{Total Wall-time}: Aggregated job wall-clock time. \\
\midrule
\textbf{Efficiency} & \textbf{CPU Utilization}: Ratio of used vs.\ allocated CPU. \\
                    & \textbf{Memory Occupancy}: Ratio of used vs.\ allocated memory. \\
                    & \textbf{Job Overhead}: Aggregated bootstrap and network-bound I/O latencies. \\
                    & \textbf{Success Rate}: Fraction of requested events completed. \\
                    & \textbf{Job Failure Rate}: Observed fraction of failed job attempts. \\
\midrule
\textbf{I/O Stress} & \textbf{Network Transfer per Event}: Remote I/O footprint per unit of work. \\
                    & \textbf{Local/Remote Ratio}: Balance of scratch vs.\ net I/O. \\
\bottomrule
\end{tabularx}
\end{table}

These metrics provide the raw data for the policy-driven score function, which allows system operators to navigate the Pareto frontier of workflow execution based on localized infrastructure priorities.

\section{Experimental Design and Scenario Modeling}
\label{sec:experimental}
To evaluate the proposed grouping strategies, we execute a multi-variate factorial experiment across diverse workflow profiles and four environmental stressors. While the framework supports complex DAGs - including fork structures - this study focuses on a rigorous characterization of three sequential profiles to achieve causal isolation. This allows for a precise analysis of the interplay between taskset granularity and resource constraints without the confounding variables of non-linear dependencies. The broader dataset, including results for an additional three fork-based profiles, is documented and available in the public repository.

\begin{figure}[b]
\centering
\includegraphics[width=\columnwidth]{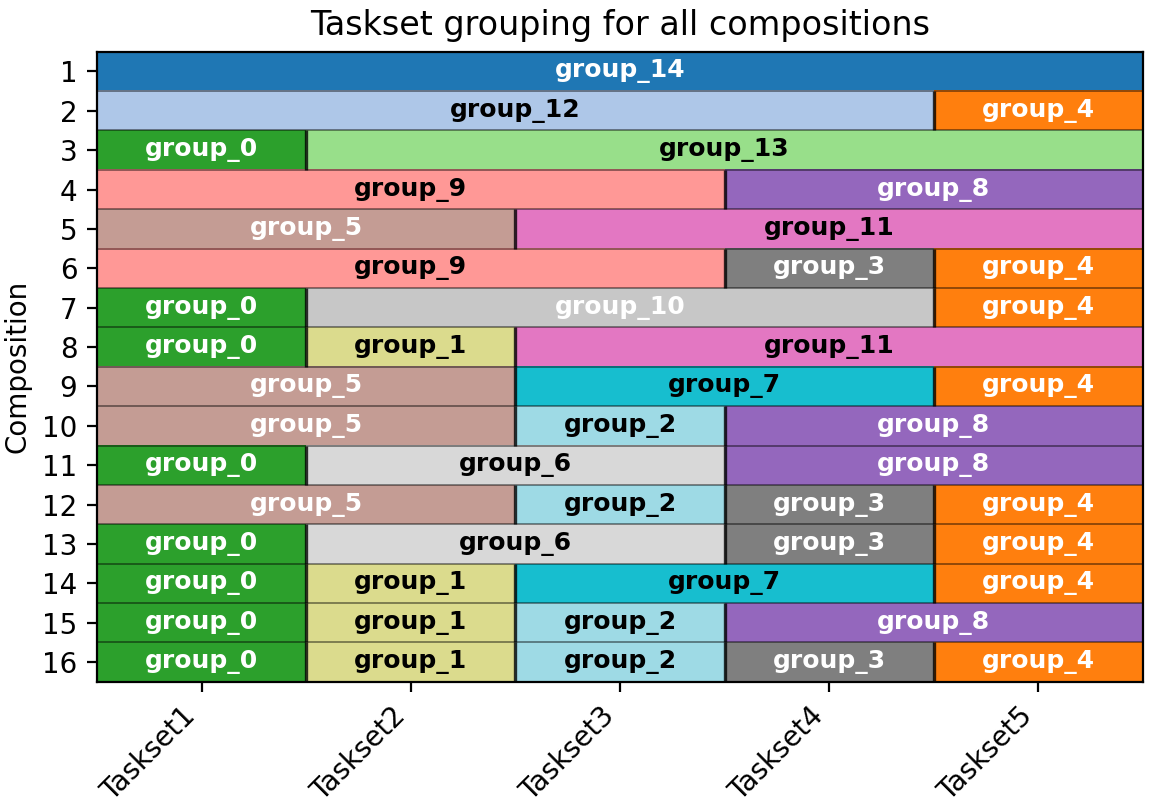}
\caption{Structural overview of the 16 evaluated workflow configurations, illustrating the spectrum from a monolithic grouped model (Const 1: StepChain) to a fully decomposed independent model (Const 16: TaskChain). Color-coded segments indicate shared execution boundaries for Tasksets 1--5 (e.g., Const 7 aggregates Tasksets 2--4 while isolating Tasksets 1 and 5).}
\label{fig:grouping-overview}
\end{figure}

\subsection{Workflow Profiles and Composition Search Space}
\label{sec:profiles}

We define three 5-taskset sequential profiles to capture the diversity of HEP workloads, with their specific resource requirements detailed in Table~\ref{tab:profile-details}. Each profile is evaluated across 16 different compositions (denotated as Const 1 through Const 16, short for Workflow Construction 1 to 16), ranging from the fully-grouped \textit{StepChain} (Const 1) to the fully-independent \textit{TaskChain} (Const 16), as illustrated in Fig.~\ref{fig:grouping-overview}.

\begin{figure}[t]
\centering
\includegraphics[width=\columnwidth]{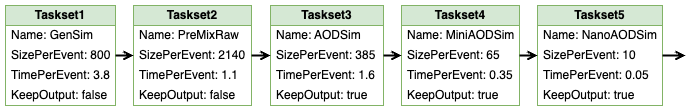}
\caption{Characterization of baseline taskset parameters across the three evaluated workflow profiles. Output data products for Taskset 1 and Taskset 2 are transient within the runtime pipeline and are not persisted for final end-user analysis.}
\label{fig:cms-seq-realistic-resources}
\end{figure}

\begin{itemize}
    \item \textbf{Realistic (\textit{seq\_real})}: A CMS-inspired sequential production chain featuring heterogeneous taskset requirements. As detailed in Fig.~\ref{fig:cms-seq-realistic-resources}, tasksets are characterized by specific $TimePerEvent$ and $SizePerEvent$ values calibrated from real-world WLCG production runs. This profile includes transient data handling for intermediate tasksets.
    \item \textbf{Homogeneous (\textit{seq\_homo})}: Features uniform resource requirements across all tasksets. This profile isolates the pure I/O and execution overhead by eliminating resource fragmentation and allocation waste.
    \item \textbf{Heterogeneous (\textit{seq\_hetero})}: Features extreme variance in requirements, specifically a 64-core requirement for the intermediate reconstruction stage (Taskset3). This profile evaluates the over-provisioning penalty: coarse-grained grouping (Const 1) forces the system to allocate peak requirements (64 cores) for the entire execution duration, leading to significant resource fragmentation.
\end{itemize}

\begin{table}[t]
\centering
\caption{Taskset Resource Requirements across Workflow Profiles.}
\label{tab:profile-details}
\small
\begin{tabularx}{\columnwidth}{@{}l ccc ccc@{}}
\toprule
 & \multicolumn{2}{c}{\textbf{seq\_real}} & \multicolumn{2}{c}{\textbf{seq\_homo}} & \multicolumn{2}{c}{\textbf{seq\_hetero}} \\
\cmidrule(lr){2-3} \cmidrule(lr){4-5} \cmidrule(lr){6-7}
\textbf{Taskset} & \textbf{Cores} & \textbf{RAM} & \textbf{Cores} & \textbf{RAM} & \textbf{Cores} & \textbf{RAM} \\
\midrule
T1 & 8 & 3GB & 8 & 8GB & 1  & 2GB  \\
T2 & 4 & 7GB & 8 & 8GB & 8  & 16GB \\
T3 & 4 & 8GB & 8 & 8GB & 64 & 64GB \\
T4 & 2 & 4GB & 8 & 8GB & 4  & 10GB \\
T5 & 4 & 4GB & 8 & 8GB & 4  & 8GB  \\
\bottomrule
\end{tabularx}
\end{table}

\subsection{Experimental Matrix}
\label{sec:matrix}
\begin{figure}[ht]
\centering
\includegraphics[width=\columnwidth]{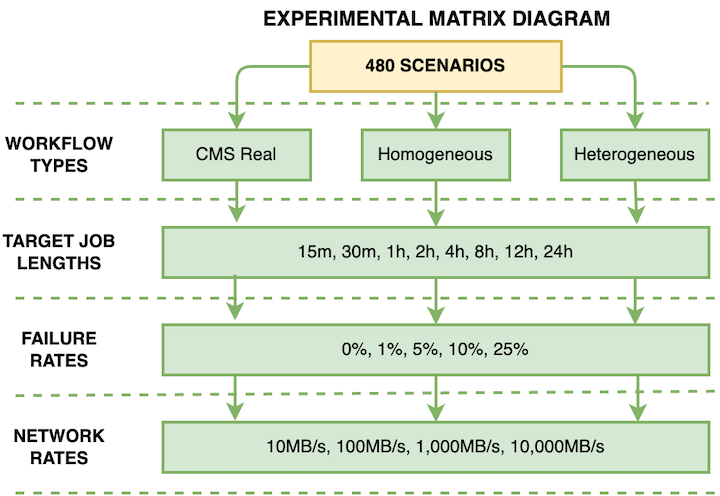}\\[0.5em]
\caption{Multi-variate experimental matrix mapping the 480 evaluation scenarios designed to expose architectural limits and systems-level scaling trade-offs.}
\label{fig:experimental-matrix}
\end{figure}

The study explores the complete combinatorial space of workflow profiles, target job lengths, failure rates and network bandwidths, resulting in 480 unique scenarios for each composition. For a sequential fully compatible 5-tasksets workflow, there are 16 valid workflow constructions, leading to a total of 7,680 unique simulation runs. Fig.~\ref{fig:experimental-matrix} summarizes these experimental factors and their systems-level rationale. To ensure a controlled environment, we assume a fixed bootstrap overhead (60s) and a saturated resource pool with unlimited job slots to focus purely on the structural efficiency of the compositions.

\subsection{Empirical Validation and Calibration}
\label{sec:calibration}
Before conducting wide-scale parameter sweeps, DAGFlowSim was validated against production WLCG execution traces using the \textit{Realistic} template. We calibrated baseline simulation parameters using aggregated metadata from actual CMS StepChain (Const 1) and TaskChain (Const 16) runs, normalized across a 5-million event envelope. These operational extrema represent the only execution modes natively supported by current CMS workload management infrastructure (WMCore); validating intermediate hybrid constructions (e.g., Const 13) on production grid infrastructure would require extensive middleware refactoring, a software development effort that this paper's simulation framework is specifically designed to motivate and inform. To isolate structural I/O dynamics from transient infrastructure noise, empirical metrics were normalized across successful job executions, imputing sample means for transient job failures. Corresponding simulations were executed under a baseline 0\% failure rate policy, yielding deterministic results with zero variance across independent runs.

\begin{figure}[t]
\centering
\includegraphics[width=\columnwidth]{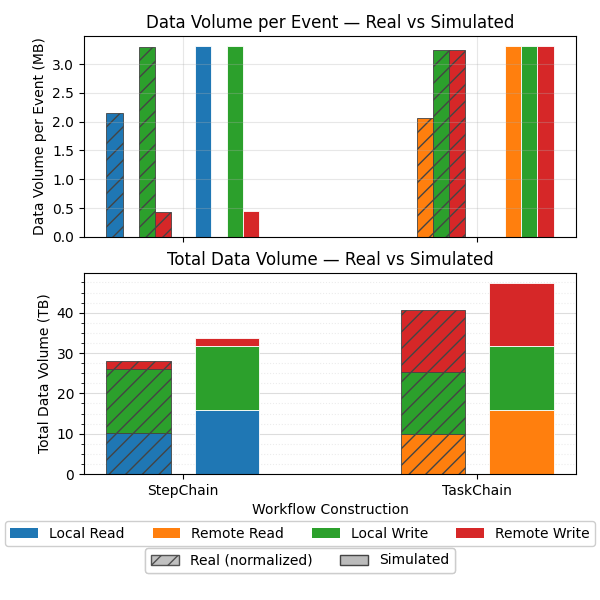}
\caption{Simulator validation against empirical WLCG baselines under zero-failure conditions (12h target length, 100~MB/s bandwidth). Hatched bars denote empirical metrics normalized across successful jobs; solid bars represent deterministic simulations (zero variance). Though absolute empirical I/O is lower due to unconsumed secondary data products, structural scaling trends and ratios between StepChain and TaskChain are accurately captured.}
\label{fig:real-vs-simulated}
\end{figure}

As shown in Fig.~\ref{fig:real-vs-simulated}, the simulator accurately replicates fundamental structural I/O trade-offs. The total absolute data volume is lower in empirical runs because production workflows generate secondary data products that are persisted but never read by downstream tasksets in the processing chain. However, the performance scaling and proportional I/O ratio between the aggregated StepChain and decomposed TaskChain models remain tight and consistent, validating the model's predictive accuracy for comparative structural analysis.

\section{Performance Evaluation and Trade-off Analysis}
\label{sec:evaluation}
We evaluate the 16 compositions across the full factorial experimental matrix to identify the Pareto optimal points within the grouping spectrum. This evaluation is structured as a three-stage analysis pipeline:

\begin{enumerate}
    \item \textbf{Explore}: A baseline characterization of the fundamental trade-offs between I/O pressure, resource efficiency, and aggregate throughput.
    \item \textbf{Generalize}: A sensitivity analysis under environmental stressors - including infrastructure faults, bandwidth constraints, and varying taskset granularities - to assess composition robustness.
    \item \textbf{Rank}: A policy-driven synthesis using a multi-metric score function to identify optimal compositions for specific operational regimes.
\end{enumerate}

By traversing this evaluation pipeline, we demostrate how the optimal grouping strategy shifts from extrema toward hybrid compositions, as environmental constraints and infrastructure volatility increase.

 \subsection{Baseline Compositional Trade-offs}
\label{sec:baseline}
We establish a baseline utilizing the \textit{seq\_real} profile (12h jobs, 100~MB/s, 0\% failure) to isolate the structural impact of composition on I/O pressure, resource fragmentation, and aggregate throughput.

\begin{figure}[t]
\centering
\includegraphics[width=\columnwidth]{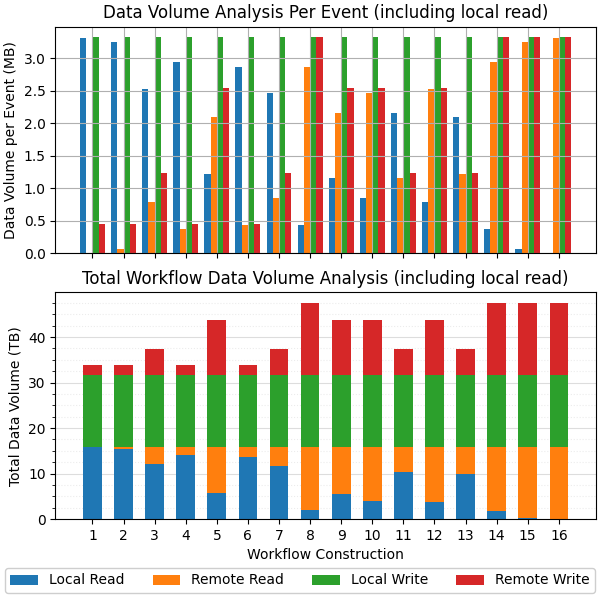}
\caption{I/O volume breakdown across the 16 compositions under baseline CMS realistic profile parameters. \textit{Top}: Per-event data volume isolated by operation type. \textit{Bottom}: Cumulative stacked workflow data volume. Grouped compositions (Const 1) optimize local data reuse, whereas full independence (Const 16) maximizes remote data-staging overhead.}
\label{fig:rq2-io}
\end{figure}

\textbf{I/O and Storage Footprint}: As shown in Fig.~\ref{fig:rq2-io}, maximal grouping (Const 1) reduces remote I/O  by $14.9\times$ compared to the full independence (Const 16) by leveraging worker-node scratch space. Conversely, fully ungrouped composition maximizes the network-transfer-per-event, as every taskset transition incurs a remote staging cost.

\begin{figure}[t]
\centering
\includegraphics[width=\columnwidth]{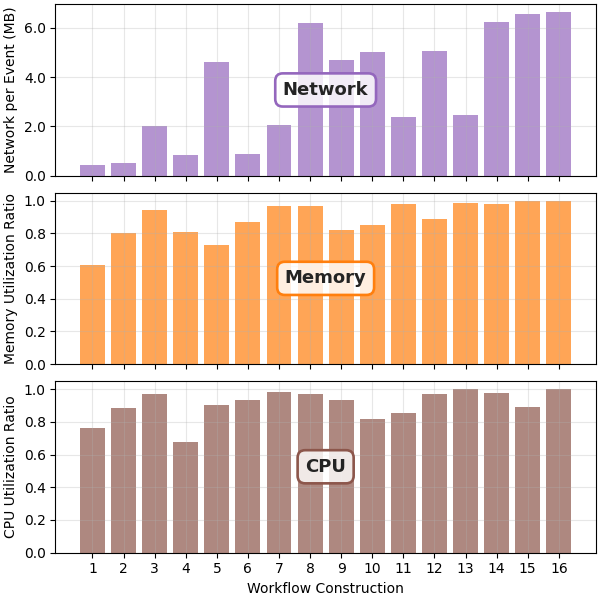}
\caption{Resource utilization profiles across all 16 compositions under the baseline scenario (\textit{seq\_real}, 12h target length, 0\% failure rate, 100~MB/s bandwidth). Monolithic grouping (Const 1) minimizes network transfer but underutilizes CPU and memory; conversely, high independence (Const 16) maximizes CPU and memory efficiency at the cost of severe network-staging overhead.}
\label{fig:rq2-resource}
\end{figure}

\begin{figure}[b]
\centering
\includegraphics[width=\columnwidth]{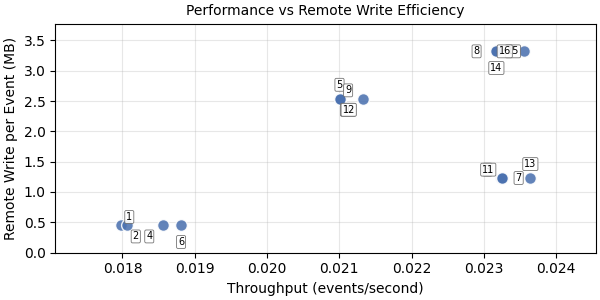}
\caption{Trade-off evaluation of the 16 compositions under the baseline scenario (\textit{seq\_real}, 12h target length, 0\% failure rate, 100~MB/s bandwidth), mapping event throughput against shared-storage write volumes. The distribution identifies an optimal composition that achieves maximum throughput while reducing remote write overhead by a factor of $\sim$3$\times$.}
\label{fig:rq2-perf}
\end{figure}

\textbf{Resource Fragmentation and Efficiency}: While Const 1 excels in I/O, it suffers from severe resource underutilization due to over-provisioning. Because a grouped job must request the maximum CPU and memory required by any single taskset for its entire duration, it leads to significant allocation waste during low-intensity tasks. Fig.~\ref{fig:rq2-resource} quantifies this fragmentation by showing the resource utilization rate differences among the configurations. Hybrid compositions (e.g., Const 13) bridge this gap, yielding a \textbf{31.5\% higher throughput} and reducing CPU and memory overhead by \textbf{23.6\% and 39.2\%}, respectively. Fig.~\ref{fig:rq2-perf} visualizes the resulting network-throughput \textbf{Pareto frontier}.

\textbf{Strategic Takeaway}: The grouping spectrum reveals a non-monotonic relationship between granularity and performance, where hybrid compositions mitigate the diminishing returns of both over-aggregation and over-decomposition.

\subsection{Sensitivity to Environmental Stressors}
\label{sec:stressors}
We evaluate the robustness of composition strategies by subjecting them to four dimensions of operational stress: \textbf{Temporal} (Job Length), \textbf{Reliability} (Failures), \textbf{Infrastructure} (Bandwidth), and \textbf{Structural} (Workflow Profiles). The "Best Hybrid" for each scenario is defined by the highest event throughput, with network transfer used as a tiebreaker.

\begin{figure}[t]
\centering
\includegraphics[width=\columnwidth]{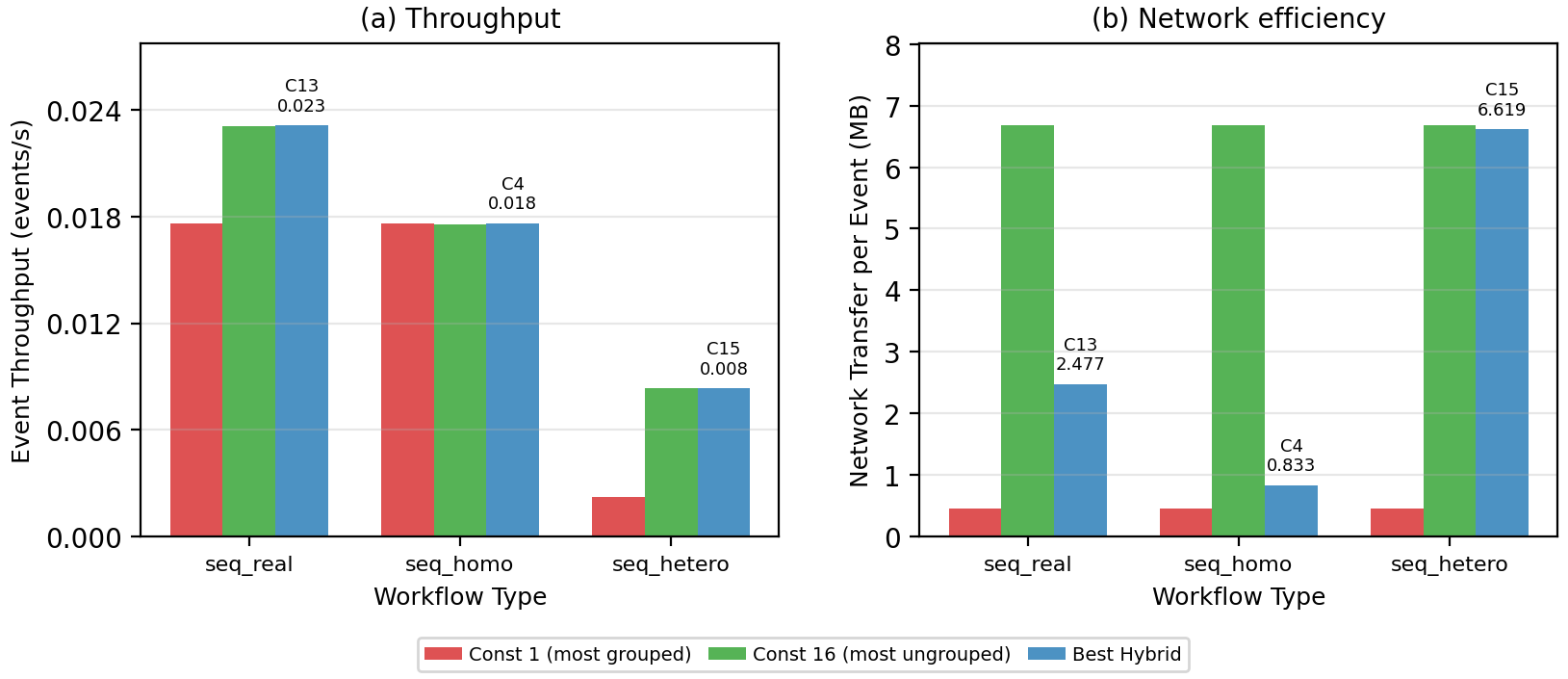}
\caption{Impact of workflow profiles on system performance (5\% failure rate, 12h target length). 
(a) Event Throughput: Const 16 provides the highest performance gain in heterogeneous workloads, while homogeneous chains exhibit minimal variance. 
(b) Network Efficiency: Const 1 minimizes per-event I/O overhead whereas Const 16 maximizes it across all workflow profiles; hybrid configurations achieve a Pareto-optimal balance.}
\label{fig:merged-workflow-sensitivity}
\end{figure}

\textbf{Structural and Infrastructure Constraints}: Workflow topology fundamentally dictates strategy success. As shown in Fig.~\ref{fig:merged-workflow-sensitivity}a, the Heterogeneous profile exhibits the highest sensitivity; grouping diverse tasksets into a monolithic unit (Const 1) results in a $3.8\times$ throughput penalty due to resource over-provisioning. Conversely, Fig.~\ref{fig:merged-workflow-sensitivity}b illustrates the I/O cost of decomposition: Const 16 incurs up to $14.9\times$ more remote network activity than Const 1. While a 10~MB/s network bandwidth severely constrains I/O-intensive tasks (causing overheads up to $\sim$4,000s), performance stabilizes at 100~MB/s as network latency ceases to be the primary bottleneck.

\textbf{Reliability and Failures}: While higher failure rates (up to 25\%) increase network waste due to retries, the throughput degradation remains largely consistent across all compositions. Notably, the hybrid Const 13 matches TaskChain throughput while achieving a \textbf{2.7$\times$ network efficiency improvement} over the fully decomposed model. This demonstrates that hybrid compositions provide superior resilience by limiting the I/O blast radius without sacrificing productivity.

\textbf{Temporal Dynamics and Makespan}: Target job length creates a sharp trade-off between amortized overhead and completion speed. Short-lived jobs (15m-1h) amplify the fixed 60s bootstrap overhead per taskset, consuming up to \textbf{33\% of total wall-clock time} for grouped units. However, makespan (Total Turnaround Time) heavily favors Const 1, which completes in 36h compared to 71h for Const 16 (at 12h length, 5\% failure). This confirms that sites prioritizing rapid completion of urgent tasks may favor StepChains, while those prioritizing global efficiency should favor hybrids.

\textbf{Strategic Takeaway}: While event throughput is resilient to failures, resource efficiency (network and overhead) is highly sensitive to the operating environment. This volatility necessitates a hybrid approach to balance the low latency of StepChains with the high resource utilization of TaskChains.

\subsection{Policy-Driven Composition Ranking}
\label{sec:policy-driven}
The final stage of our evaluation synthesizes multi-dimensional metrics into an actionable utility ranking. We define a weighted score function $S$ for each composition $C$:
\begin{equation}
S(C) = \sum_{i} (w_i \times \hat{m}_i)
\label{eq:score}
\end{equation}
where $\hat{m}_i$ is the normalized, feature-scaled value of metric $i$ (throughput, CPU/event, Mem/event, and Network/event) such that higher values denote more desirable performance, and $w_i$ represents the policy-specific weights ($\sum w_i = 1$).

\begin{figure}[t]
\centering
\includegraphics[width=\columnwidth]{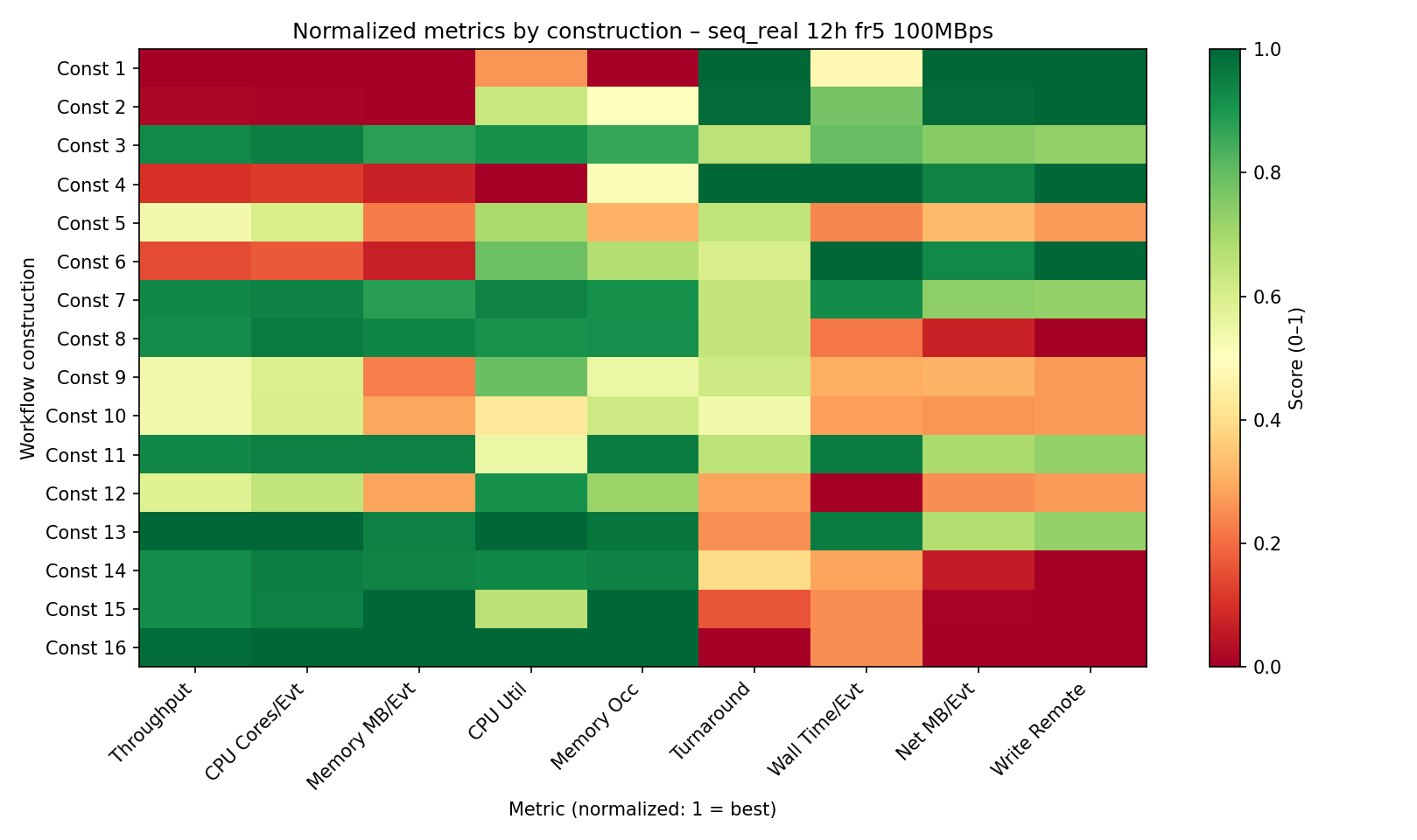}
\caption{Heatmap of normalized metric scores for all 16 compositions under the realistic sequential workflow (5\% failure rate, 12h target length, 100~MB/s bandwidth). Highly grouped (e.g., Const 1) compositions excel at I/O efficiency while high independence (e.g., Const 16) optimizes throughput and resource utilization; hybrid configurations successfully balance both performance extremes.}
\label{fig:heatmap}
\end{figure}

The metric heatmap in Fig.~\ref{fig:heatmap} provides a global visualization of this design space. The two extrema - the fully aggregated \textit{StepChain} (Const 1) and the fully independent \textit{TaskChain} (Const 16) - occupy opposite ends of the spectrum, each excelling in one dimension while sacrificing others. Specifically, highly independent compositions optimize throughput and resource utilization but exhibit poor performance in makespan and remote I/O efficiency. Conversely, highly aggregated compositions lead in I/O efficiency but lack throughput and resource-allocation efficiency. Hybrid constructions, particularly Const 13, emerge as Pareto-efficient configurations, capturing high scores across multiple dimensions and avoiding the catastrophic inefficiencies of the extreme policies.

To demonstrate the framework's flexibility, we evaluate three optimization policies (Table~\ref{tab:score-policies}) tailored to distinct operational and infrastructure constraints. The \textit{Default Policy} prioritizes throughput while balancing resource consumption, whereas the \textit{I/O-Prioritized} and \textit{Resource-Prioritized} policies shift the focus to specific system bottlenecks.

\begin{table}[ht]
\centering
\caption{Optimization Policy Weight Configurations}
\label{tab:score-policies}
\small
\begin{tabularx}{\columnwidth}{@{}l cccc X@{}}
\toprule
\textbf{Policy} & \textbf{Thr.} & \textbf{CPU} & \textbf{Mem} & \textbf{Net.} & \textbf{Primary Objective} \\
\midrule
Default & 0.4 & 0.2 & 0.2 & 0.2 & Balance throughput and efficiency. \\
\addlinespace[2pt]
I/O-Prior. & 0.2 & 0.2 & 0.2 & 0.4 & Minimize shared storage footprint. \\
\addlinespace[2pt]
Res.-Prior. & 0.2 & 0.3 & 0.3 & 0.2 & Maximize per-event utilization. \\
\bottomrule
\end{tabularx}
\end{table}

\begin{figure}[t]
\centering
\includegraphics[width=\columnwidth]{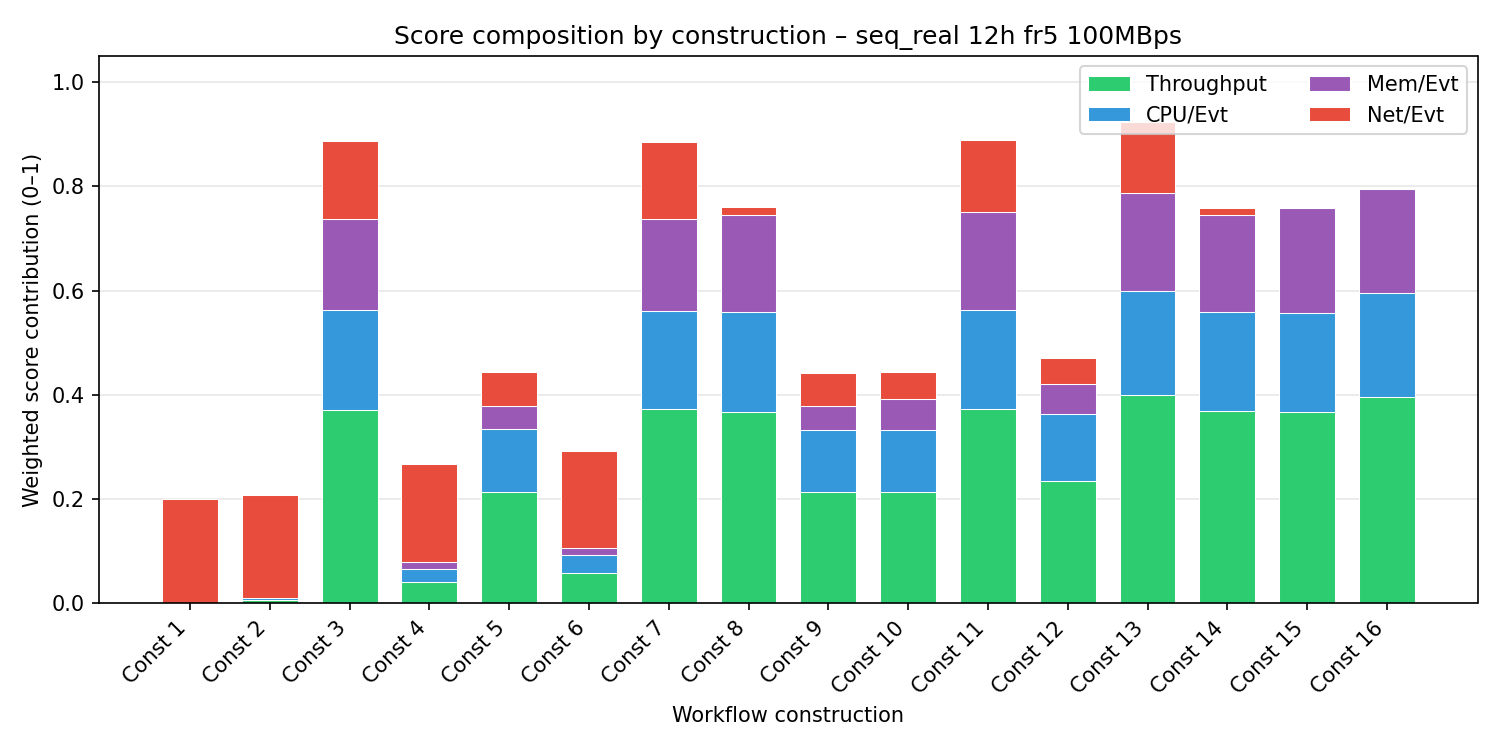}
\caption{Workflow composition score breakdown under the \textit{Default Policy} (maximizing throughput). The chart illustrates the cumulative contribution of each evaluation metric to the final score. Although Const 13 and Const 16 both maximize throughput (yielding the full 0.4 score component), the network footprint acts as the decisive differentiator favoring the hybrid configuration strategy.}
\label{fig:stacked_score}
\end{figure}

In our realistic case study, Const 13 remains the top-ranked configuration across all three policies. It successfully mitigates the static resource fragmentation of Const 1 while bypassing the severe I/O penalties of Const 16. Fig.~\ref{fig:stacked_score} visualizes this behavior under the \textit{Default Policy}, showing the breakdown of how each metric's weighted contribution ($w_i \times \hat{m}_i$) accumulates toward a perfect utility score of 1.0. While both Const 13 and Const 16 maximize the throughput component (capturing the full 0.4 weight allocation), Const 16 suffers a heavy network penalty. This makes network efficiency the decisive factor that establishes the hybrid Const 13 as the globally optimal construction.

Our results confirm that the optimal composition is highly sensitive to the underlying workload profile. In the \textbf{Homogeneous} profile, the StepChain becomes competitive (Score: 0.74) as the over-provisioning penalty is eliminated. However, in the \textbf{Heterogeneous} profile, aggressive ungrouping (Const 15 and 16) becomes mandatory to avoid resource waste. This confirms that a static, invariant grouping strategy is fundamentally flawed for modern, heterogeneous grid computing; instead, composition must be a policy-driven, workload-aware decision.

\section{Discussion and Recommendations}
\label{sec:discussion}
The simulation results validate our central thesis: optimal workflow compositions are rarely found at the static extrema (\textit{StepChain} or \textit{TaskChain}) but rather in hybrid configurations that balance I/O locality with resource throughput.

Analyzing the \textit{seq\_real} profile under the \textit{Default Policy}, Const 13 (a \textbf{1+2+1+1} grouping) emerges as the Pareto-optimal construction for this specific operational scenario. We emphasize that Const 13 is an \textit{exemplar output} of our policy-driven framework rather than a universal static recipe for all HEP workloads. In this specific 5-taskset workflow, Const 13 targets the most I/O-intensive segment: Tasksets 2 and 3 are grouped to keep heavy intermediate data local to the worker node scratch disk, while Tasksets 1, 4, and 5 execute independently. This eliminates a major remote I/O bottleneck without incurring the severe resource over-provisioning penalties of a full StepChain.

\subsection{Practical Recommendations for Workflow Engineering}
\label{sec:recommendations}
To assist practitioners in navigating this trade-off space, we synthesize our findings into a clear decision logic based on infrastructure and workflow-specific constraints:

\begin{itemize}
    \item \textbf{For Network-Constrained Sites ($<100$~MB/s)}: High-granularity compositions (e.g., TaskChain) incur prohibitive network overhead. Designers should favor locality-heavy hybrids (e.g., Const 3 or 7), which offer superior throughput compared to a StepChain while maintaining a minimal remote I/O footprint.    
    \item \textbf{For Heterogeneous Workflows}: When tasksets exhibit high variance in resource requirements, functional decomposition is mandatory. Since a StepChain approach incurs a $3.8\times$ throughput penalty in such scenarios, designers should use utilization-optimized hybrids (e.g., Const 13-16) to enable fine-grained resource allocation.    
    \item \textbf{For Time-Critical Campaigns}: If makespan is prioritized over global efficiency, StepChains (Const 1) remain superior for highly concurrent resource pools, as they eliminate inter-job scheduling latencies and data-staging delays.
\end{itemize}

\subsection{Scope and Limitations}
\label{sec:limitations}
While \textit{DAGFlowSim} is calibrated against real WLCG production runs, it is designed for comparative trade-off analysis rather than absolute metric prediction. Current limitations include a simplified failure model omitting site-correlated outages and an assumption of worker homogeneity. Nevertheless, relative performance rankings and the identification of Pareto-efficient hybrid compositions remain robust across these operational variables.

Furthermore, specific compositional outcomes, such as Const 13 emerging as optimal in the \textit{seq\_real} baseline, are strictly instance-dependent, governed by the DAG topology, taskset resource variance, and localized policy weights ($w_i$). As demonstrated by our \textit{seq\_homo} and \textit{seq\_hetero} evaluations, changing the underlying taskset requirements shifts the Pareto frontier, rendering more aggregated or decomposed configurations optimal. The primary contribution of this work is therefore not any single static construction index, but rather the optimization framework itself, which consistently identifies the superior hybrid operating point for a given workflow and site profile.

\section{Conclusion and Future Work}
\label{sec:conclusion}
This paper introduces a formal framework for taskset composition alongside a multi-metric score function to automate extreme-scale scientific workflow orchestration. By exploring the spectrum between traditional grouped (\textit{StepChain}) and fully decomposed (\textit{TaskChain}) models, we demonstrate that workload-aware hybrid workflow compositions can match the throughput of independent tasks while significantly mitigating the I/O burden on shared infrastructure. Ultimately, this framework provides the High-Energy Physics and broader distributed computing communities with a quantitative, policy-driven methodology to dynamically identify optimal configurations and balance the competing demands of data locality and resource efficiency.

\textbf{Future Work}: We plan to extend this framework for the HL-LHC era across four primary vectors. For \textbf{Topological Complexity}, we will support multi-parent DAG structures (e.g., diamond or join shapes) to accommodate complex physics processing chains. For \textbf{Hardware Heterogeneity}, we will integrate support for heterogeneous accelerators (GPUs/FPGAs) and opportunistic HPC resources, where resource over-provisioning is heavily penalized. For \textbf{I/O Characterization}, we plan to enrich workflow descriptions to account for unconsumed secondary data products, tracking the precise ratio between stored and active downstream inputs per event. Finally, for \textbf{WMS Integration}, we plan to integrate a lightweight framework version into the future CMS Workload Management System to enable autonomic workflow composition based on site policies.

\bibliographystyle{IEEEtran}
\bibliography{biblio}

\end{document}